\documentclass[aps,pra,reprint,amsmath,amssymb,superscriptaddress,nofootinbib,longbibliography]{revtex4-2}

\usepackage{graphicx}
\usepackage[utf8]{inputenc}
\usepackage[T1]{fontenc}
\usepackage{xcolor}
\usepackage{soul}

\usepackage{amssymb}
\usepackage{bm}

\begin{document}

%\begin{frontmatter}


%% \title{Title\tnoteref{}}

\title{Crystal and magnetic structures of $R_2$Ni$_2$In compounds ($R$ = Tb and Ho)}


%\author[IFJ]{Aleksandra Deptuch}
%\author[CUT]{Ryszard Duraj}
%\author[IFUJ]{Andrzej Szytu\l{}a}
%\author[IFUJ]{Bogus\l{}aw Penc}
%\author[IFJ,OU]{Ewa Juszy\'nska-Ga\l{}\k{a}zka}
%\author[IFUJ]{Stanis\l{}aw Baran\corref{cor1}}

%\address[IFJ]{Institute of Nuclear Physics Polish Academy of 
%Sciences, Radzikowskiego 152, PL-31-342 Krak\'ow, Poland}
%\address[CUT]{Institute of Physics, Cracow University of Technology, Podchor\k{a}\.zych 1,
%PL-30-084 Krak\'ow, Poland}
%\address[IFUJ]{M.~Smoluchowski Institute of Physics, Jagiellonian
%University, prof. Stanis\l{}awa \L{}ojasiewicza 11, PL-30-348 Krak\'ow,
%Poland}
%\address[OU]{Department of Chemistry, Graduate School of Science, Osaka University,
%1-1 Machikaneyama, Toyonaka, 560-0043, Osaka, Japan}
%\cortext[cor1]{Corresponding author. Phone: (+48) 126644686;
%e-mail: \mbox{stanislaw.baran@uj.edu.pl}}

\author{Stanis\l{}aw Baran}
\email{stanislaw.baran@uj.edu.pl}
\affiliation{M.~Smoluchowski Institute of Physics, Jagiellonian University,
prof. Stanis\l{}awa \L{}ojasiewicza 11, PL-30-348 Krak\'ow, Poland}
\author{Aleksandra Deptuch}
\affiliation{Institute of Nuclear Physics Polish Academy of
Sciences, Radzikowskiego 152, PL-31-342 Krak\'ow, Poland}
\author{Andreas Hoser}
\affiliation{Helmholtz-Zentrum Berlin f\"ur Materialien und Energie GmbH, 
Hahn-Meitner Platz~1, D-14109, Berlin, Germany}
\author{Bogus\l{}aw Penc}
\affiliation{M.~Smoluchowski Institute of Physics, Jagiellonian University,
prof. Stanis\l{}awa \L{}ojasiewicza 11, PL-30-348 Krak\'ow, Poland}
\author{Janusz Przewo\'znik}
\affiliation{AGH University of Science and Technology, Faculty of Physics and Applied Computer Science,
Department of Solid State Physics, Al. Mickiewicza 30, PL-30-059 Krak\'ow, Poland}
%
%\author{Yuriy Tyvanchuk}
%\affiliation{Analytical Chemistry Department, Ivan Franko National University of Lviv,
%Kyryla i Mefodiya 6, 79005 Lviv, Ukraine}
%
\author{Andrzej Szytu\l{}a}
\affiliation{M.~Smoluchowski Institute of Physics, Jagiellonian University,
prof. Stanis\l{}awa \L{}ojasiewicza 11, PL-30-348 Krak\'ow, Poland}

\date{\today}

\begin{abstract}

Crystal and magnetic structures of $R_2$Ni$_2$In ($R$ = Tb and Ho) have been studied by powder neutron
diffraction at low temperatures. The compounds crystallize in an orthorhombic crystal structure of the
Mn$_2$AlB$_2$-type. At low temperatures, the magnetic moments localized solely on the rare earth atoms
form antiferromagnetic structures. The Tb magnetic moments, equal to 8.65(6) $\mu_B$ and
parallel to the $c$-axis, form a collinear magnetic structure described by the propagation vector
$\boldsymbol{k} = [\frac{1}{2}, \frac{1}{2}, \frac{1}{2}]$. This magnetic structure is stable up to the
N\'eel temperature equal to 40~K. For Ho$_2$Ni$_2$In a complex, temperature-dependent
magnetic structure is detected. In the temperature range 3.5-8.6~K, an incommensurate magnetic
structure, described by the propagation vector $\boldsymbol{k}_1 = [0.76, 0, 0.52]$ is observed, while in the
temperature interval 2.2-3.1~K the magnetic order is described by two propagation vectors, namely,
$\boldsymbol{k}_2 = [\frac{5}{6}, 0.16, \frac{1}{2}]$ and its third harmonics $3\boldsymbol{k}_2 = [\frac{5}{2}, 0.48, \frac{3}{2}]$.
Below 2~K, a coexistence of all magnetic structures detected at higher temperatures is observed.
For all magnetic phases, the Ho magnetic moments are parallel to the $c$-axis. The low temperature
heat capacity data confirm a first order transition near 3~K.

\bigskip

%\noindent \textbf{keywords}: intermetallic compounds, X-ray diffraction (XRD), ferromagnetic materials,
%magnetic properties, pressure-temperature phase diagram

\end{abstract}

\maketitle

%\begin{keyword}

%x-ray diffraction, ac magnetic susceptibility, pressure, (P,T) phase diagram, magnetostructural coupling

%intermetallic compounds \sep X-ray diffraction (XRD) \sep ferromagnetic materials \sep
%magnetic properties \sep pressure-temperature phase diagram

%%Intermetallics \sep Magnetic properties \sep Ferromagnetic \sep 
%%Pressure-temperature phase diagram \sep X-ray diffraction (XRD)


%\PACS 61.12.-q \sep  75.25.+z \sep 75.50.Ee


%\end{keyword}

%\setcounter{page}{0}

%\end{frontmatter}

% Commented as '\usepackage{lineno}' has been commented too above
%\linenumbers


\section{Introduction}
\label{intro}

There is a strong interest in the development of novel magnetically ordered materials with unconventional properties.
The $R_2$Ni$_2$In compounds crystallize in the orthorhombic crystal structure (space group $Cmmm$)
\cite{Zaremba_Izv_Akad_Nauk_24}. Magnetic and specific heat measurements indicate that the $R_2$Ni$_2$In ($R$ = Gd-Tm)
compounds are antiferromagnets with the N\'eel temperatures between 5~K ($R$ = Tm) and 40~K ($R$ = Tb)
\cite{Szytula_JMMM_387}. In all compounds, except $R$ = Ho, the magnetic order is stable in a broad
temperature range between 1.9~K and a corresponding N\'eel temperature. For Ho$_2$Ni$_2$In, below the N\'eel temperature
of 9~K, an additional phase transition at $T_t$ equal 3.5 K (dc magnetic data) or 3.3~K (ac
magnetic data and heat capacity data) is observed. The more recent paper reports
$T_N = 10.5$~K and $T_t = 5.5$~K~\cite{Zhang_Dalton_Trans_48}. The antiferromagnetic order is confirmed by the neutron diffraction data for
$R$ = Tb~\cite{Szytula_Acta_Phys_Pol_A_124} as well as Er and Tm~\cite{Baran_JAC_696}. The magnetic structure is
described by the propagation vectors $\boldsymbol{k} = [\frac{1}{2}, \frac{1}{2}, \frac{1}{2}]$ for $R$ = Tb and
$[\frac{1}{2}, 0, \frac{1}{2}]$ for Er and Tm. The magnetic moments are found to be localized solely on the rare earth
atoms. They form collinear magnetic structure and are parallel to the $c$-axis for Tb and
$b$-axis for Er and Tm.

In order to deeper understand magnetic properties of $R_2$Ni$_2$In ($R$ - rare earth element), we have
performed new neutron diffraction measurements for $R_2$Ni$_2$In ($R$ = Tb, Ho). Although, the magnetic structure of
Tb$_2$Ni$_2$In has already been reported~\cite{Szytula_Acta_Phys_Pol_A_124}, the previous data have been collected 
for the sample containing only 10~wt~\% of Tb$_2$Ni$_2$In. In the current work we report the results obtained for a new sample
consisting of Tb$_2$Ni$_2$In in its most part. In addition, we report for the first time the magnetic structure in Ho$_2$Ni$_2$In.
The neutron diffraction data for Ho$_2$Ni$_2$In are supported by heat capacity measurements.

\section{Experimental details}

The samples of $R_2$Ni$_2$In ($R$ = Tb and Ho) have been prepared by arc melting of $R$, Ni and In
(all with purity at least 99.9~wt~\%) taken in the atomic ratio of 2:2:1. The melting
has been performed under a Ti-gettered Ar atmosphere. The obtained ingots have been turned over and remelted four
times in order to get homogeneous distribution of components. The samples have been homogenized in an evacuated
quartz-tube at 873~K for one month, followed by cold water quenching.

The samples' quality has been checked by X-ray powder diffraction at room temperature (PANalytical X'Pert diffractometer with CuK$\alpha$ radiation).

Neutron diffraction patterns have been collected in the temperature range from 1.55 up to 60~K
on the E6 diffractometer at the Helmholtz-Zentrum Berlin f\"ur Materialien in Energie GmbH. The incident neutron wavelength was 2.4315~\AA{}.
For Rietveld analysis of the X-ray and neutron diffraction patterns the computer program FullProf has been utilized \cite{Rodriguez-Carvajal_Newsletter_26},
while for symmetry analysis the computer program BasIreps, which is distributed together with FullProf, has been used.

Heat capacity study have been carried out by a two-tau relaxation method in the temperature range 1.8-12~K using HC option
of the Quantum Design PPMS platform.

\section{Results}

\begin{figure}[!ht]
\begin{center}
\includegraphics[bb=14 14 559 808, width=\columnwidth]
        {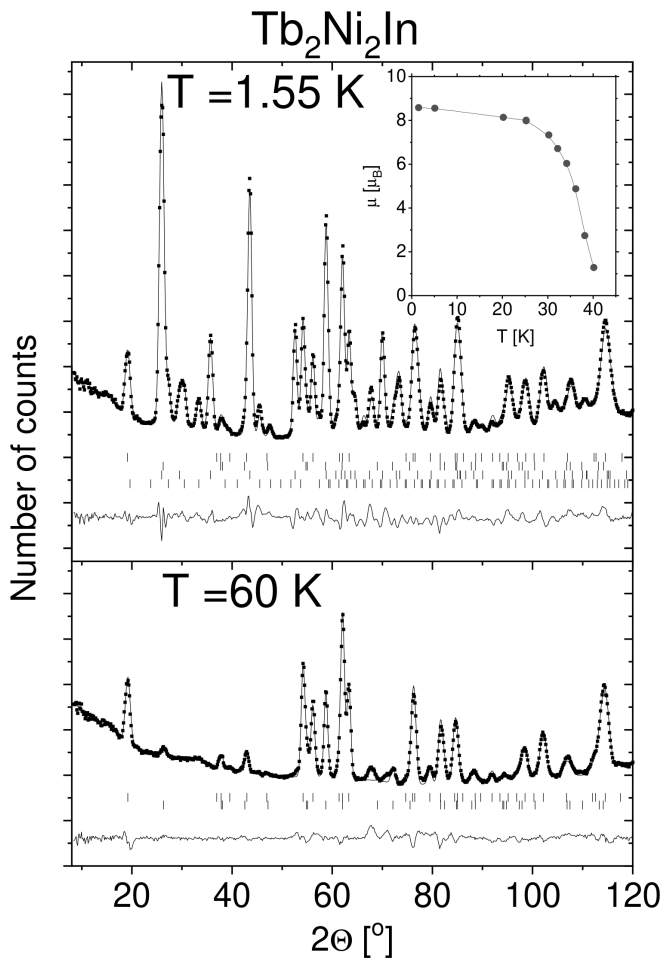}
\end{center}
\caption{\label{fig:Tbneutron}
Neutron diffraction patterns of Tb$_2$Ni$_2$In collected at 1.55 and 60~K. In both patterns the squares
represent experimental points. The solid lines are the calculated profiles for the crystal and magnetic 
structure models (as is described in text) and difference between the observed and calculated intensities (at the
bottom of each diagram). The vertical bars indicate the positions of nuclear (first top and third rows) and magnetic
(second and fourth row) peaks for Tb$_2$Ni$_2$In and Tb$_2$Ni$_{1.78}$In, respectively. The insert shows the
temperature dependence of the Tb magnetic moment.}
\end{figure}

\begin{figure*}[!ht]
\begin{center}
\includegraphics[bb=14 14 2510 1489, width=\textwidth]
        {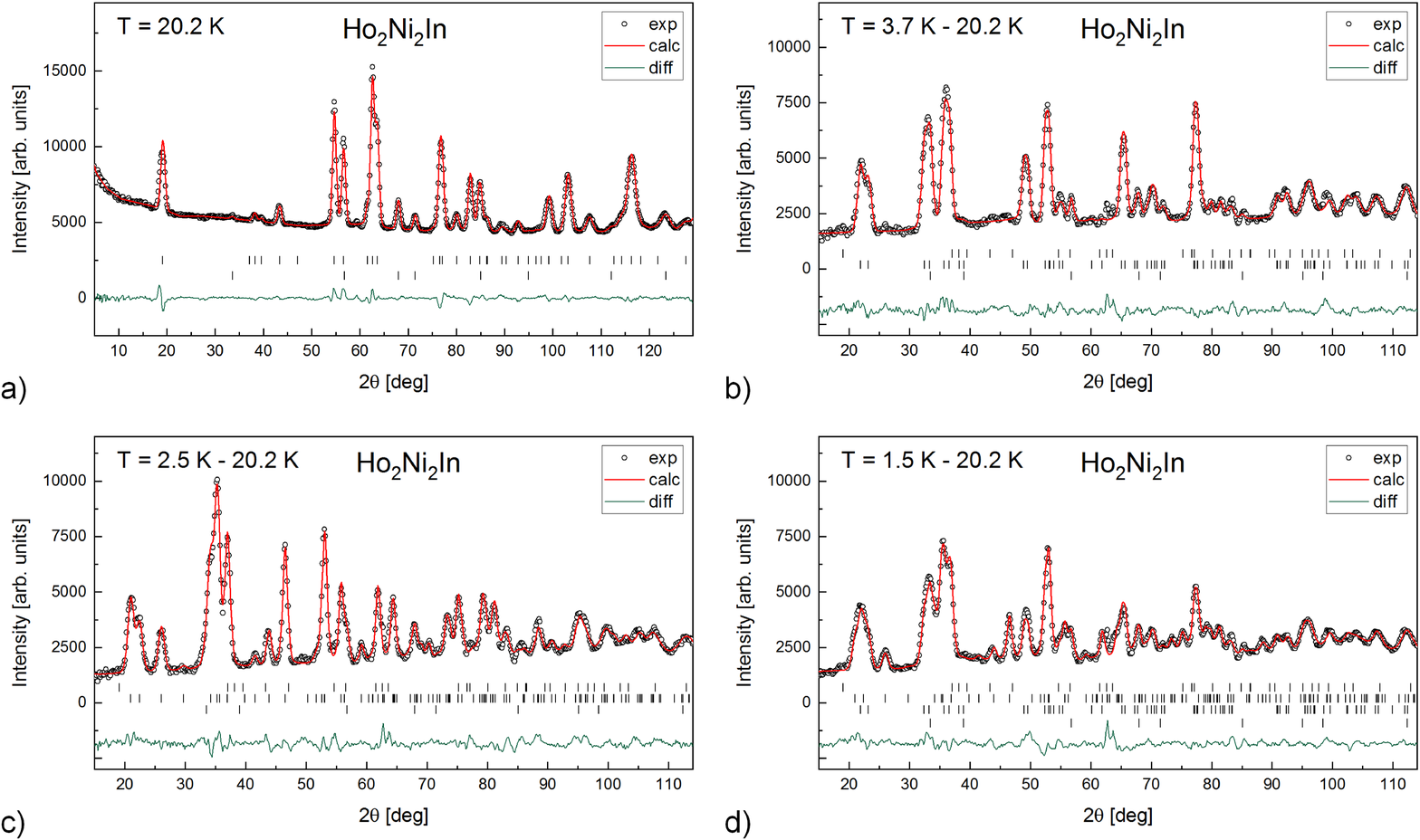}
\end{center}
\caption{\label{fig:Honeutron}
Neutron diffraction patterns of Ho$_2$Ni$_2$In collected at a) $T$ = 20.2 K, b) $T$ = 3.7 K, c) $T$ = 2.5 K and
d) $T$ = 1.5 K. Experimental patterns are denoted by points, while the results of Rietveld refinement by
lines and the difference curves by bottom lines. From patterns b)-d), the pattern taken at
$T = 20.2$~K has been subtracted in order to obtain pure magnetic contribution.
The first row of vertical bars indicates Bragg reflection positions originating from the crystal structure of Ho$_2$Ni$_2$In.
The magnetic Bragg reflection positions of Ho$_2$Ni$_2$In related to the propagation vector $\boldsymbol{k}_1 = [0.76, 0, 0.52]$
are denoted by the second row in b) and third row in d). The magnetic Bragg reflection positions of Ho$_2$Ni$_2$In corresponding to the propagation vectors 
$\boldsymbol{k}_2 = [\frac{5}{6}, 0.16, \frac{1}{2}]$ and $3\boldsymbol{k}_2 = [\frac{5}{2}, 0.48, \frac{3}{2}]$ are denoted by the second row in c) and d).
The nuclear peak positions of HoNi$_2$ impurity are denoted by the second row in a). 
The magnetic peaks from ferromagnetic order in HoNi$_2$ are indicated by the last rows in b)-d).}
\end{figure*}

Analysis of the X-ray diffraction and neutron patterns indicate 82~wt~\% of Tb$_2$Ni$_2$In and 18~wt~\% of Tb$_2$Ni$_{1.78}$In
for the Tb-based sample and 95.4(3)~wt~\% of Ho$_2$Ni$_2$In and 4.6(3)~wt~\% of HoNi$_2$ for the
Ho-based sample (see Figures~\ref{fig:Tbneutron}a and \ref{fig:Honeutron}a).

\begin{figure}[!ht]
\begin{center}
\includegraphics[bb=14 14 444 592, width=0.5\columnwidth]
        {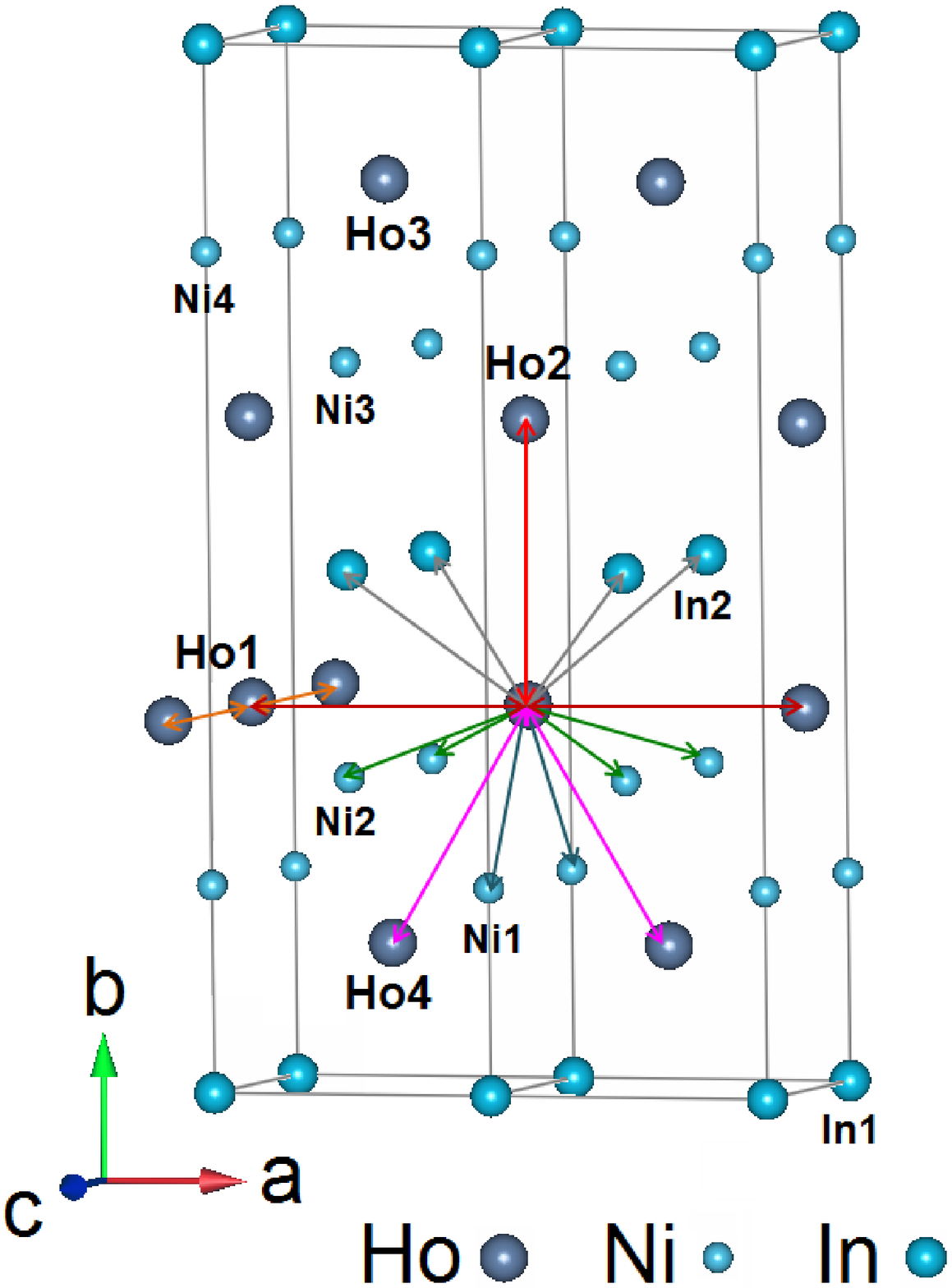}
\end{center}
\caption{\label{fig:structure}
Crystal unit cell of Ho$_2$Ni$_2$In (doubled along the $a$-axis for better visibility). The distances between the Ho1 atom and its nearest neighbours are indicated.}
\end{figure}

\begin{table}
\caption{\label{tab:Tbcryst}
Crystal structure parameters of Tb$_2$Ni$_2$In together with residuals for profile and integrated intensities determined from the neutron diffraction data.}
%\begin{footnotesize}
%\begin{tabular*}{0.7\textwidth}{c@{\extracolsep{\fill}}cc}
\begin{tabular*}{0.7\columnwidth}{c@{\extracolsep{\fill}}cc}
%\begin{tabular}{c@{\extracolsep{\fill}}cc}
\hline
$T$ [K] & 60 & 1.55 \\ \hline
$a$ [\AA{}] & 3.913(1) & 3.910(1) \\
$b$ [\AA{}] & 14.135(2) & 14.125(3) \\
$c$ [\AA{}] & 3.691(1) & 3.692(1) \\
$V$ [\AA{}$^3$] & 204.15(14) & 203.90(16) \\
$y_{Tb}$ & 0.3647(2) & 0.3632(3) \\
$y_{Ni}$ & 0.1984(3) & 0.1986(3) \\ \hline
$R_{Bragg}$ [\%] & 6.2 & 6.6 \\
$R_f$ [\%] & 5.0 & 5.8 \\ \hline
\end{tabular*}
%\end{tabular}
%\end{footnotesize}
\end{table}

\begin{table}
\begin{center}
\caption{\label{tab:Hocryst}
Crystal structure parameters of Ho$_2$Ni$_2$In together with residuals for profile and integrated intensities determined from the neutron diffraction data.}
%\begin{tabular*}{1\textwidth}{c@{\extracolsep{\fill}}cccc}
\begin{tabular*}{1\columnwidth}{c@{\extracolsep{\fill}}cccc}
\hline
$T$ [K] & 20.2 & 3.7 & 2.5 & 1.5 \\ \hline
$a$ [\AA{}] & 3.9178(5) & 3.9119(9) & 3.9117(8) & 3.9122(9) \\
$b$ [\AA{}] & 14.202(2) & 14.184(3) & 14.184(3) & 14.184(3) \\
$c$ [\AA{}] & 3.6715(4) & 3.6648(8) & 3.6660(7) & 3.6652(8) \\
$V$ [\AA{}$^3$] & 204.29(2) & 203.34(8) & 203.40(7) & 203.38(7) \\
$y_{Ho}$ & 0.3639(4) & 0.3636(6) & 0.3639(6) & 0.3644(6) \\
$y_{Ni}$ & 0.1981(4)& 0.1972(8) & 0.1969(8) & 0.1975(7) \\ \hline
$R_{Bragg}$ [\%] & 3.4 & 5.1 & 4.9 & 5.3 \\
$R_f$ [\%] & 2.8 & 3.6 & 3.0 & 3.2 \\ \hline
\end{tabular*}
\end{center}
\end{table}

\begin{table}
\begin{center}
\caption{\label{tab:HoNN}
Interatomic distances between the Ho1 atom and its nearest neighbours in Ho$_2$Ni$_2$In determined from the full diffraction patterns.}
%\begin{tabular*}{1\textwidth}{c@{\extracolsep{\fill}}cccc}
\begin{tabular*}{1\columnwidth}{c@{\extracolsep{\fill}}cccc}
\hline
distance [\AA{}] & 20.2 K & 3.7 K & 2.5 K & 1.5 K \\ \hline
Ho1-Ni2 & 2.825(3) & 2.815(5) & 2.816(5) & 2.821(4) \\
Ho1-Ni1 & 2.986(6) & 2.99(2) & 3.00(1) & 2.99(1) \\
Ho1-In1 & 3.308(3) & 3.306(5) & 3.303(5) & 3.299(5) \\
Ho1-Ho1 & 3.6715(4) & 3.6648(8) & 3.6660(7) & 3.6652(8) \\
Ho1-Ho4 & 3.782(7) & 3.77(1) & 3.78(1) & 3.79(1) \\
Ho1-Ho2 & 3.866(8) & 3.87(2) & 3.86(1) & 3.85(2) \\ \hline
\end{tabular*}
\end{center}
\end{table}

The 2:2:1 compounds crystalize in the orthorhombic crystal structure of the Mn$_2$AlB$_2$-type (space group $Cmmm$). The atoms occupy the following Wyckoff sites:
$R$ at $4j$ $(0, y_R, \frac{1}{2})$, Ni at $4i$ $(0, y_{Ni}, 0)$ and In at $2a$ $(0, 0, 0)$. The crystal structure of Ho$_2$Ni$_2$In
is shown in Figure~\ref{fig:structure} together with distances between the Ho1 atom and its nearest
neighbours. The refined crystal structure parameters at different temperatures are listed in Table~\ref{tab:Tbcryst} for
Tb$_2$Ni$_2$In and Table~\ref{tab:Hocryst} for Ho$_2$Ni$_2$In, respectively. The data confirm that the crystal 
structure is stable down to low temperatures. The shortest interatomic distances between the Ho1 atom
and its nearest neighbours in Ho$_2$Ni$_2$In are listed in Table~\ref{tab:HoNN}.

Four magnetic $R$ atoms in the $4j$ sublattice with the local symmetry $m2m$ are denoted as: $R1$ $(0, y_R, \frac{1}{2})$,
$R2$ $(0, 1-y_R, \frac{1}{2})$, $R3$ $(\frac{1}{2}, y_R+\frac{1}{2}, \frac{1}{2})$, $R4$ $(\frac{1}{2}, \frac{1}{2}-y_R, \frac{1}{2})$.
The $R3$ and $R4$ atom positions are related to the $R1$ and $R2$ ones, respectively, by the centering
translation $C = [\frac{1}{2}, \frac{1}{2}, 0]$.

\begin{figure}[!ht]
\begin{center}
\includegraphics[bb=14 14 252 549, width=0.5\columnwidth]
        {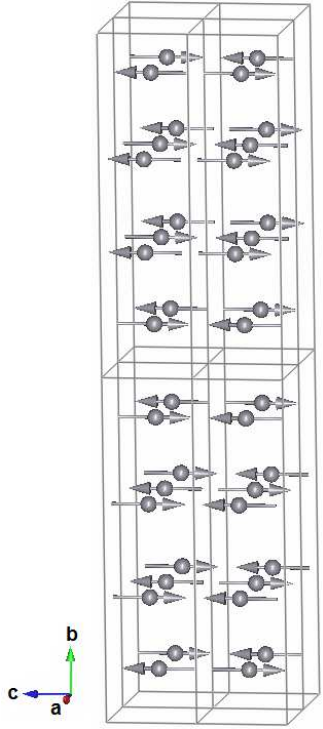}
\end{center}
\caption{\label{fig:Tbcell}
Magnetic unit cell of Tb$_2$Ni$_2$In. It is doubled along three directions when compared with the crystallographic one.}
\end{figure}

The neutron diffraction pattern of Tb$_2$Ni$_2$In, collected at 1.55~K (see Figure~\ref{fig:Tbneutron}b), contains additional Bragg
reflections originating from the magnetic order. These reflections can be indexed with the use of the propagation vector
$\boldsymbol{k} = [\frac{1}{2}, \frac{1}{2}, \frac{1}{2}]$. The same propagation vector describes magnetic order
in the isostructural Er$_2$Ni$_2$Pb compound \cite{Prokes_PRB_78}. Symmetry analysis provides four magnetic structure models
related to $\boldsymbol{k}$ -- for details see Table II in~\cite{Prokes_PRB_78}.
In case of Tb$_2$Ni$_2$In, the best agreement with the experimental data ($R_{magn}$ = 5.0\%) is obtained for the model
$\Gamma 3$ with Tb magnetic moments equal to 8.65(6)~$\mu_B$ and parallel to the $c$-axis. The moments follow the $++-+$ sign sequence
for the Tb1, Tb2, Tb3 and Tb4 atoms in the crystallographic unit cell, respectively. The corresponding magnetic unit cell is shown in
Figure~\ref{fig:Tbcell}. Temperature dependence of the Tb magnetic moment (see the insert in Figure~\ref{fig:Tbneutron})
provides the N\'eel temperature equal to 40~K, which is in a good agreement with the results of macroscopic 
magnetic measurements~\cite{Szytula_JMMM_387}.

\begin{figure}[!ht]
\begin{center}
\includegraphics[bb=14 14 564 698, width=1.0\columnwidth]
        {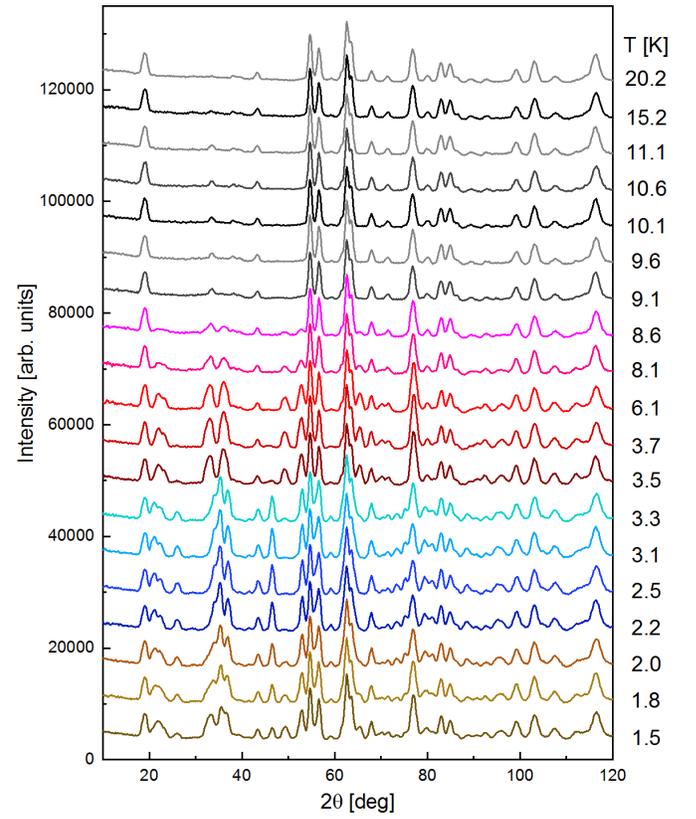}
\end{center}
\caption{\label{fig:Hotemp}
Thermal evolution of neutron diffraction patterns of Ho$_2$Ni$_2$In.}
\end{figure}

\begin{figure}[!ht]
\begin{center}
\includegraphics[bb=14 14 1581 880, width=1.0\columnwidth]
        {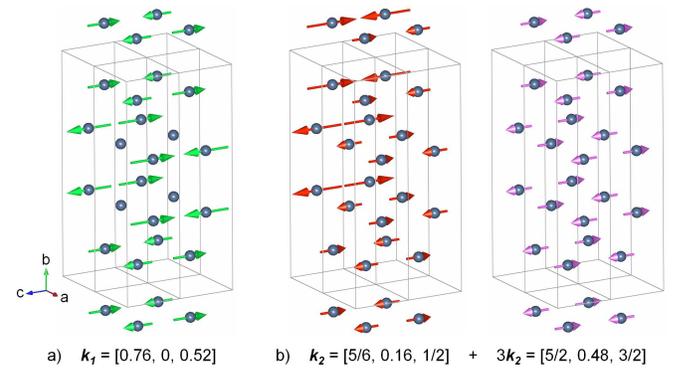}
\end{center}
\caption{\label{fig:Homagns}
Modulated magnetic structures of Ho$_2$Ni$_2$In described by the propagation vectors: a) $\boldsymbol{k}_1 = [0.76, 0, 0.52]$ at $T$ = 3.7 K and
b) $\boldsymbol{k}_2 = [\frac{5}{6}, 0.16, \frac{1}{2}]$ and $3\boldsymbol{k}_2 = [\frac{5}{2}, 0.48, \frac{3}{2}]$ at $T$ = 2.5 K.}
\end{figure}

Neutron diffraction patterns of Ho$_2$Ni$_2$In, collected in the 1.5-20.2~K range, are presented in Figure~\ref{fig:Hotemp}.
%, were analyzed by the Rietveld refinement in FullProf program \cite{Rodriguez-Carvajal_Newsletter_26}.
Ho$_2$Ni$_2$In remains paramagnetic down to 9.1~K. The Bragg reflections of magnetic origin appear in the pattern collected
at 8.6~K. With decreasing temperature the Ho$_2$Ni$_2$In magnetic structure undergoes two transitions visible as distinct changes
in the diffraction pattern at ca. 3.5~K and ca. 2.0~K, respectively. In order to extract pure magnetic contribution, the paramagnetic
pattern taken at $T=20.2$~K has been subtracted from the patterns recorded at lower temperatures -- representative examples are shown
in Figures~\ref{fig:Honeutron}b-d. Propagation vectors describing magnetic structure have been determined using the k-search computer
program, while for symmetry analysis of magnetic structures the BasIreps program has been utilized. Both programs are parts of the FullProf
package~\cite{Rodriguez-Carvajal_Newsletter_26}. The ferromagnetic contribution arising from the HoNi$_2$ impurity phase has been included
in all refinements performed for data collected below 13.5~K~\cite{Cwik_JAC_735}. It has been assumed that all magnetic moments in HoNi$_2$
are of the same magnitude and point at the [111] direction.

%Down to 9.1 K, 
%Ho$_2$Ni$_2$In is in a paramagnetic state. The magnetic peaks appear in 8.6 K and as it can be seen in Figure~\ref{Hotemp}, down to 1.5 K Ho$_2$Ni$_2$In undergoes two magnetic phase transitions, one at ca. 3.5 K and another one at 
%ca. 2.0 K. The propagation vectors and possible irreducible representations of magnetic structures for 4j Wyckoff site were obtained in k-search and BasIreps programs, respectively, both from FullProf package. The values of magnetic 
%moments were determined by the Rietveld refinement from difference patterns containing only magnetic peaks, obtained by subtraction of the pattern registered in $T$ = 20.2 K from the patterns collected below the magnetic ordering 
%temperature (representative refinements presented in Figures~\ref{Honeutron}b-d). The contribution from ferromagnetic order in HoNi$_2$ was included below 13.5 K \cite{Cwik_JAC_735}, with assumption of equal magnetic 
%moments ordered 
%along [111] direction.

\begin{table*}
\begin{center}
\caption{\label{tab:Homagn1}
Magnetic structure of Ho$_2$Ni$_2$In described by the IR2 representation for the $4j$ Wyckoff site of $Cmmm$ space group
and propagation vector $\boldsymbol{k}_1 = [k_x, 0, k_z]$. $C_i$ coefficients denote contribution to the total magnetic moment
along respective BVi basis vector, $\mu_{tot}$ is the modulation amplitude of magnetic moment and $R_{magn}$ is the magnetic
reliability factor of the Rietveld refinement. Only the values for Ho1 and Ho2 atoms are given since Ho3 and Ho4 
atoms are interrelated with them by centering translation.}
%\begin{tabular*}{1\textwidth}{c@{\extracolsep{\fill}}ccccccccccc}
\begin{tabular*}{1\textwidth}{c@{\extracolsep{\fill}}ccccccccccc}
%\begin{tabular*}{1\columnwidth}{c@{\extracolsep{\fill}}cccc}
\hline
 & & & IR1 & & & IR2 \\
%\hline
\cline{3-5} \cline{6-8}
 & $\boldsymbol{k}_1 = [k_x, 0, k_z]$ & BV1 & BV2 & BV3 & BV1 & BV2 & BV3 \\
\hline
$4j$ & Ho1 \\
 & $(0, y_{Ho}, \frac{1}{2})$ & 1 0 0 & 0 1 0 & 0 0 1 & 1 0 0 & 0 1 0 & 0 0 1 \\
 & Ho2 \\
  & $(0, 1-y_{Ho}, \frac{1}{2})$ & $-$1 0 0 & 0 1 0 & 0 0 $-$1 & 1 0 0 & 0 $-$1 0 & 0 0 1 \\ \hline
$T$ & Ho$_2$Ni$_2$In & $C_1$ & $C_1$ & $C_1$ & $C_1$ & $C_1$ & $C_1$ & $\mu_{tot}$ & $k_x$ & $k_z$ & $R_{magn}$ \\
~[K] & & [$\mu_B$] & [$\mu_B$] & [$\mu_B$] & [$\mu_B$] & [$\mu_B$] & [$\mu_B$] & [$\mu_B$] & & & [\%] \\ \hline
3.7 & Ho1 & & & & & & 10.28(9) & 10.28(9) & 0.7580(3) & 0.5198(3) & 5.3 \\
 & Ho2 & & & & & & 10.28(9) & 10.28(9) \\
1.5 & Ho1 & & & & & & 8.2(2) & 8.2(2) & 0.7568(5) & 0.5191(5) & 5.4 \\
 & Ho2 & & & & & & 8.2(2) & 8.2(2) \\ \hline
\end{tabular*}
\end{center}
\end{table*}

\begin{figure}[!ht]
\begin{center}
\includegraphics[bb=14 14 696 549, width=1.0\columnwidth]
        {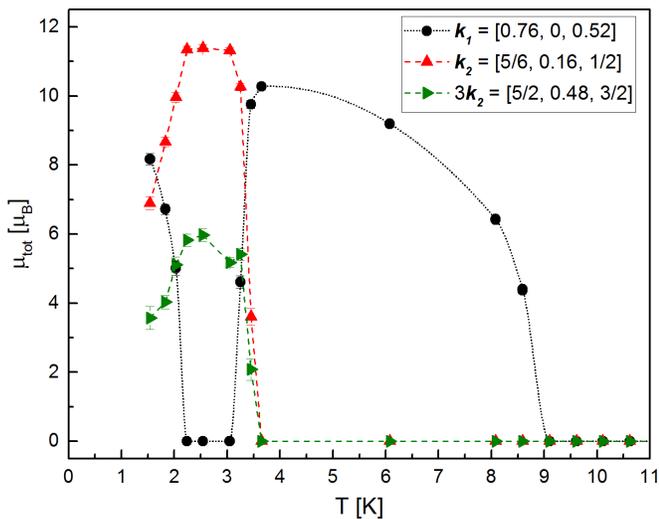}
\end{center}
\caption{\label{fig:Homagn}
Modulation amplitude of the Ho magnetic moments in Ho$_2$Ni$_2$In in function of temperature as obtained by the Rietveld refinement
from difference diffraction patterns (representative fits shown in Figures~\ref{fig:Honeutron}b-d). The modulation is described either
by the propagation vector $\boldsymbol{k}_1 = [0.76, 0, 0.52]$ or $\boldsymbol{k}_2 = [\frac{5}{6}, 0.16, \frac{1}{2}]$
and its third harmonics $3\boldsymbol{k}_2 = [\frac{5}{2}, 0.48, \frac{3}{2}]$. The lines connecting 
points are only guides to eye.}
\end{figure}

The Bragg reflections of magnetic origin, observed in the 3.5-8.6~K temperature range, can be indexed with an incommensurate propagation vector
$\boldsymbol{k}_1 = [0.76, 0, 0.52]$. Symmetry analysis for $\boldsymbol{k}_1$ and the $4j$ Wyckoff site provides two
irreducible representations: IR1 and IR2. Both representations allow for non-zero contribution to the total magnetic moments along
$a$, $b$, $c$ directions (Table~\ref{tab:Homagn1}). All Ho atoms belong to one orbit, therefore they are constrained by symmetry and
modulation of their magnetic moments has identical amplitude. The IR1 representation assumes parallel ordering of the
$b$-axis components of magnetic structure and antiparallel ordering of the $a$- and $c$-axis components for the Ho1 and Ho2 atoms belonging
to the same crystal unit cell. For IR2, the $b$-axis components are antiparallel, while the $a$- and $c$-axis ones are parallel. The results
of Rietveld refinement favor the magnetic structure model related to the IR2 representation with magnetic moments oriented only along
the $c$-axis. A representative difference pattern ($T = 3.7$~K - 20.2~K) is presented in Figure~\ref{fig:Honeutron}b, while the 
corresponding magnetic structure is shown in Figure~\ref{fig:Homagns}a. The resultant magnetic structure is antiferromagnetic.
The modulation amplitude increases initially with decreasing temperature reaching 10.3(1)~$\mu_B$ at $T$~=~3.7~K
(see Figure~\ref{fig:Homagn}). At the same temperature, new unindexed reflections of low intensity become visible, indicating
beginning of magnetic phase transition. At $T$ = 3.5~K and 3.3~K, a co-existence of two different magnetic phases is observed.
The modulation amplitude related to the $\boldsymbol{k}_1 = [0.76, 0, 0.52]$ propagation vector decreases to 9.8(2)~$\mu_B$ and
4.6(2)~$\mu_B$ at 3.5~K and 3.3~K, respectively, and finally drops to zero at $T$ = 3.1 K.

\begin{table*}
%\begin{tiny}
\begin{center}
\caption{\label{tab:Homagn2}
Magnetic structure of Ho$_2$Ni$_2$In described by the IR2 representation for the $4j$ Wyckoff site of $Cmmm$ space group
and propagation vectors $\boldsymbol{k}_2 = [\frac{5}{6}, 0.16, \frac{1}{2}]$ and
$3\boldsymbol{k}_2 = [\frac{5}{2}, 0.48, \frac{3}{2}]$.
$C_1$ coefficient is a contribution to the magnetic moment along BV1 basis vector (the first and second values apply to
$\boldsymbol{k}_2$ and $3\boldsymbol{k}_2$, respectively), $\mu_{tot}$ is the modulation amplitude of magnetic moment, while
$R_{magn}$ is the magnetic reliability factor of Rietveld refinement.}
%\begin{tabular*}{1\textwidth}{c@{\extracolsep{\fill}}ccccccccccc}
\begin{tabular*}{1\textwidth}{c@{\extracolsep{\fill}}ccccccc}
%\begin{tabular*}{1\columnwidth}{c@{\extracolsep{\fill}}cccc}
\hline
%& & & IR1 & & IR2 \\ \hline
 & & \multicolumn{2}{c}{IR1} & IR2 \\
%\hline
\cline{3-4} \cline{5-5}
 & $\boldsymbol{k}_2 = [\frac{5}{6}, 0.16, \frac{1}{2}]$ & BV1 & BV2 & BV1 \\
 & $3\boldsymbol{k}_2 = [\frac{5}{2}, 0.48, \frac{3}{2}]$ \\
\hline
4j & Ho1 $(0, y_{Ho}, \frac{1}{2})$ & 1 0 0 & 0 1 0 & 0 0 1 \\
 & Ho2 $(0, 1-y_{Ho}, \frac{1}{2})$ & 1 0 0 & 0 1 0 & 0 0 1 \\ \hline
$T$ [K] & Ho$_2$Ni$_2$In & $C_1$ [$\mu_B$] & $C_2$ [$\mu_B$] & $C_1$ [$\mu_B$] & $\mu_{tot}$ [$\mu_B$] & $k_y$ & $R_{magn}$ [\%] \\ \hline
2.5 & Ho1 & & & 11.41(9), 6.0(2) & 11.41(9), 6.0(2) & 0.1615 & 4.9 \\
 & Ho2 & & & 11.41(9), 6.0(2) & 11.41(9), 6.0(2) \\ \hline
1.5 & Ho1 & & & 6.9(2), 3.5(3) & 6.9(2), 3.5(3) & 0.159(2) & 5.9 \\
 & Ho2 & & & 6.9(2), 3.5(3) & 6.9(2), 3.5(3) \\ \hline
\end{tabular*}
\end{center}
%\end{tiny}
\end{table*}

Indexing of Bragg reflections of magnetic origin, observed between 2.2 and 3.1~K (see Figure~\ref{fig:Honeutron}c), requires two propagation vectors, namely,
$\boldsymbol{k}_2 = [\frac{5}{6}, 0.16, \frac{1}{2}]$ and its third harmonics, $3\boldsymbol{k}_2 = [\frac{5}{2}, 0.48, \frac{3}{2}]$,
indicating that the resultant modulation of magnetic moments differs from purely sinusoidal one, observed at higher
temperatures. Symmetry analysis provides identical irreducible representations (IR1 and IR2) for both $\boldsymbol{k}_2$ and $3\boldsymbol{k}_2$.
IR1 describes magnetic ordering within the $ab$-plane, while IR2 the one along the $c$-axis (see Table~\ref{tab:Homagn2}).
The Ho atoms are divided into two orbits -- the first one containing the Ho1, Ho3 pair and the second one containing the Ho2, Ho4 pair.
The agreement with experimental pattern is obtained for the IR2 representation (the Ho magnetic moments are parallel to the $c$-axis).
The components of magnetic structure described by the $\boldsymbol{k}_2$ propagation vector and its third harmonics $3\boldsymbol{k}_2$
are visualized in Figure~\ref{fig:Homagns}b for $T = 2.5$~K. Although the splitting into two orbits allows for different modulation amplitudes
and a phase shift between orbits, the Rietveld refinement implies identical modulation amplitudes and zero phase shift between the (Ho1, Ho3) and
(Ho2, Ho4) atom pairs. The amplitude of modulation related to the $\boldsymbol{k}_2$ propagation vector equals 11.3-11.4~$\mu_B$ in the 2.2-3.1~K
range, while the component related to $3\boldsymbol{k}_2$ is about two times smaller and equal to 5.2-6.0~$\mu_B$ (see Figure~\ref{fig:Homagn}).
This relationship between components continues also below 2.2~K, where another magnetic phase transition occurs. It is worth noting that the
magnetic structure related to $\boldsymbol{k}_2$ and $3\boldsymbol{k}_2$ resembles the high-temperature magnetic structure related to $\boldsymbol{k}_1$
as both magnetic structures are antiferromagnetic and the magnetic moments are oriented along the $c$-axis.

At the lowest studied temperatures, a~re-appearance of Bragg reflections of magnetic origin, indexed by the $\boldsymbol{k}_1 = [0.76, 0, 0.52]$
propagation vector is observed -- notice the presence of magnetic Bragg reflection at $2\theta$ = 49.3$^\circ$ in Figures~\ref{fig:Honeutron}b-d
as well as follow thermal evolution of the Ho$_2$Ni$_2$In neutron diffraction pattern shown in Figure~\ref{fig:Hotemp}.
A clear co-existence of the magnetic phase described by $\boldsymbol{k}_1$ and that related to the $\boldsymbol{k}_2$, $3\boldsymbol{k}_2$ pair
is visible below 2.0~K. The modulation amplitude of magnetic moments related to the $\boldsymbol{k}_1$ propagation vector increases with decreasing temperature
reaching finally 8.2(2)~$\mu_B$ at $T = 1.5$~K. It exceeds the contributions related to the $\boldsymbol{k}_2$ and $3\boldsymbol{k}_2$ propagation
vectors, which decrease at $T = 1.5$~K to 6.9(2)~$\mu_B$ and 3.6(4)~$\mu_B$, respectively.

\begin{figure}[!ht]
\begin{center}
\includegraphics[bb=14 14 744 547, width=\columnwidth]
        {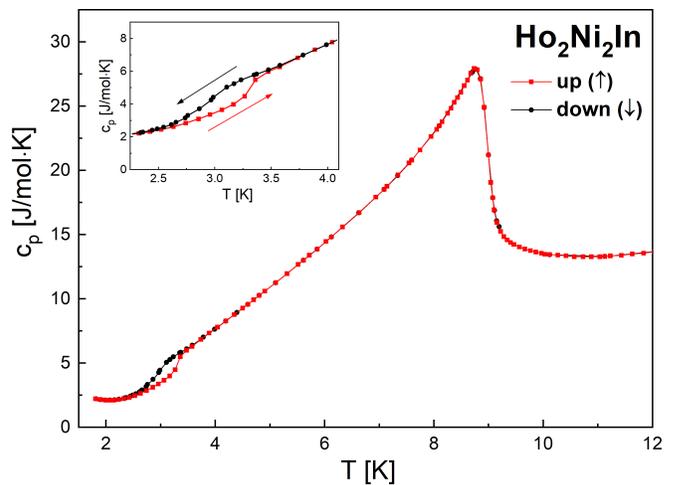}
\end{center}
\caption{\label{fig:heat}
Temperature dependence of the molar heat capacity of Ho$_2$Ni$_2$In in the temperature range 1.8-12.0~K.
The insert shows the low-temperature data with increasing (up) an decreasing (down) temperature.
The lines are guides for the eye.}
\end{figure}

In order to get more information about the nature of low-temperature magnetic phase transitions, additional heat capacity
measurements have been performed in the temperature range 1.8-12~K with both increasing and decreasing temperature (see Figure~\ref{fig:heat}).
A lambda-shaped maximum at 9.0~K is typical of the second-order transition from antiferro- to paramagnetic state, while
a~distinct thermal hysteresis visible in the temperature range 2.5-3.5~K (see the insert in Figure~\ref{fig:heat})
is characteristic of the first-order transition and coincides with the transition temperature between the magnetic phase described by
the $\boldsymbol{k}_1$ propagation vector and that related to the $\boldsymbol{k}_2$ and $3\boldsymbol{k}_2$ vector pair.

\section{Discussion}

Analysis of the X-ray diffraction pattern as well as nuclear contribution to the neutron diffraction pattern in function of temperature
indicates that the orthorhombic crystal structure is stable in a broad temperature range including the magnetically ordered state (see
Tables~\ref{tab:Tbcryst} and \ref{tab:Hocryst}). The crystal structure is highly anisotropic with the $b$ lattice constant more than three times
larger than the $a$ and $c$ ones (see Figure~\ref{fig:structure}). The structure consists of the $ac$-planes of various kinds stacked along the $b$-axis direction, namely,
the $R$ atoms planes are intercalated between the Ni- and In-planes following the sequence: In-R-Ni-Ni-R-In-R-Ni-Ni-R-In.
The shortest distances between the Ho atoms within the $ac$-plane and those located at different planes differ (see Table~\ref{tab:HoNN})
-- the lowest are found along the $c$-axis which is the direction of magnetic moment in the ordered state.

The results reported in this work indicate different kinds of collinear antiferromagnetic orderings present
in the investigated compounds, namely, a commensurate one in Tb$_2$Ni$_2$In and an incommensurate one in Ho$_2$Ni$_2$In.
In Tb$_2$Ni$_2$In, the low-temperature magnetic structure is described by the propagation vector $\boldsymbol{k} = [\frac{1}{2}, \frac{1}{2}, \frac{1}{2}]$,
leading to a magnetic unit cell that has the lattice parameters doubled with respect to the crystallographic cell along all three
principal crystallographic directions (see Figure~\ref{fig:Tbcell}). In Ho$_2$Ni$_2$In, a sequence of incommensurate
modulated magnetic structures appears with decreasing temperature. Below 9~K a magnetic structure related to $\boldsymbol{k}_1 = [0.76, 0, 0.52]$
is observed, while in the temperature range 2.2 and 3.1~K two propagation vectors ($\boldsymbol{k}_2 = [\frac{5}{6}, 0.16, \frac{1}{2}]$
and $3\boldsymbol{k}_2$) are required to describe the magnetic structure. Finally, below 2~K a~coexistence of both above mentioned
magnetic structures is detected.

%a different magnetic order in the investigated compounds: collinear in Tb$_2$Ni$_2$In and modulated in Ho$_2$Ni$_2$In. Neutron diffraction data indicate for Tb$_2$Ni$_2$In the 
%antiferromagnetic ordering described by the propagation vector $\boldsymbol{k} = [\frac{1}{2}, \frac{1}{2}, \frac{1}{2}]$ which leads to a magnetic unit cell that has the lattice parameters doubled with respect to the 
%crystallographic 
%cell along all three principal directions (see Figure~\ref{Tbcell}). For Ho$_2$Ni$_2$In the temperature dependence of the amplitude of the Ho magnetic moment (see Figure~\ref{Homagn}) gives the magnetic phase transition from para- to 
%antiferromagnetic state near 9 K are observed. Below 9 K there are three regions with different magnetic ordering with different propagation vectors: $\boldsymbol{k}_1$ in the temperature range 3.5-8.6 K with the incommensurate 
%magnetic ordering in $ac$-plane, $\boldsymbol{k}_2$ and $3\boldsymbol{k}_2$ between 2.2 and 3.1 K with the commensurate one in $ac$-plane and incommensurate along the $c$-axis, and reentrant $\boldsymbol{k}_1$ below 2 K which 
%coexists 
%with $\boldsymbol{k}_2$ and $3\boldsymbol{k}_2$.

Temperature-induced transformations of magnetic structures have been found in a number of intermetallic compounds
(see Table~1 in~\cite{Gignoux_PRB_48}). Such transformations of the magnetic structure can be understood on the basis of
a~realistic mean-field model, which takes into account both periodic-change-field and crystal electric field
effects~\cite{Gignoux_PRB_48}. A sequence of magnetic structures similar to those reported in this work for
Ho$_2$Ni$_2$In has been previously observed for UNi$_2$Si$_2$~\cite{PhysRevB.43.13232}. With decreasing temperature,
an~incommensurate modulated magnetic structure ($\boldsymbol{k} = [0,0,0.745]$) transforms into an intermediate simple
antiferromagnetic one, and finally again to a~modulated one ($\boldsymbol{k} = [0,0,\frac{2}{3}]$)
with a ferromagnetic component. Such an~evolution of magnetic structure has been interpreted on the basis of
the Heisenberg model with biquadratic exchange~\cite{PhysRevB.45.10399}.

The magnetic moments in Tb$_2$Ni$_2$In and Ho$_2$Ni$_2$In are parallel to the $c$-axis, while they are parallel to
the $b$-axis in $R_2$Ni$_2$In ($R$ = Er and Tm)~\cite{Baran_JAC_696}. Such a change of orientation of the magnetic moment
with increasing number of the $4f$ electrons can be attributed to change of sign of the Stevens operator $\alpha_J$ which is negative
for Tb and Ho and positive for Er and Tm~\cite{Stevens_Proc_Phys_Soc_A_65}.

In both compound the $R$-$R$ interatomic distances are about 3.7~\AA{} and therefore they are larger than the sum of respective
$R^{3+}$ ionic radii. Such a result suggest presence of indirect exchange interactions of the RKKY-type. The RKKY model predicts
proportionality between the N\'eel temperature and the de Gennes factor defined as $(g_J-1)^2J(J+1)$, where $g_J$ is a Land\'e
splitting factor and $J$ is a total angular momentum of the corresponding magnetic ion~\cite{Gennes_JPR_23}. Figure~11 in~\cite{Szytula_JMMM_387}
shows a comparison between the experimentally determined N\'eel temperatures for $R_2$Ni$_2$In ($R$ = Gd--Tm)
and those calculated according the RKKY theory. A large discrepancy between the experimental and calculated temperatures for Tb$_2$Ni$_2$In
indicates a strong influence of the crystalline electric field (CEF) on magnetic state formation~\cite{Noakes_PLA_91}.
Therefore the magnetic structures in $R_2$Ni$_2$In ($R$ - rare earth element) result from competition between the RKKY-
and CEF-type interactions. Such a competition may lead to complex magnetic properties including temperature-induced
magnetic order-order transitions~\cite{Rossat-Mignod_1987}. It is worth noting that such transitions
have been observed also in the isostructural $R_2$Ni$_2$Pb ($R$ = Ho, Er) compounds~\cite{Prokes_EPJB_43,Prokes_PRB_78}.
In Ho$_2$Ni$_2$Pb, a~sine-modulated commensurate magnetic order described by the magnetic unit cell $5a \times b \times c$
develops below $T_N = 7$~K. The magnetic structure turns into square-modulated one below $T_t = 3$~K. In Er$_2$Ni$_2$Pb,
an incommensurate modulated magnetic structure ($\boldsymbol{k} = [0.8409(1), 0, \frac{1}{2}]$) has been detected below $T_N = 3.5$~K. With decreasing temperature,
two intermediate incommensurate magnetic phases, related to the $[0.5973(1),0,\frac{1}{2}]$ and $[0.5330(3),0,\frac{1}{2}]$
propagation vectors, respectively, appears. A commensurate magnetic structure, involving two propagation vectors
($[\frac{1}{2},\frac{1}{2},\frac{1}{2}]$ and $[0,0,\frac{1}{2}]$), is found as magnetic ground state.

%\section{Summary}
%\section{Conclusions}

\section{Summary and conclusions}

The results presented in this work confirm that the crystal structure of ternary $R_2$Ni$_2$In ($R$ = Tb and Ho)
is orthorhombic of the Mn$_2$AlB$_2$-type (space group $Cmmm$) in both paramagnetic and magnetically ordered states.
At low temperatures the rare earth magnetic moments are found to order antiferromagnetically
with the rare earth magnetic moments being parallel to the $c$-axis. In Tb$_2$Ni$_2$In,
a~commensurate ($\boldsymbol{k}=[\frac{1}{2},\frac{1}{2},\frac{1}{2}]$) collinear antiferromagnetic structure
with is formed below $T_N = 40$~K. In Ho$_2$Ni$_2$In, a sequence of order-order magnetic transitions is observed.
An incommensurate antiferromagnetic sine-modulated structure, related to $\boldsymbol{k}_1 = [0.76, 0, 0.52]$, is detected
below $T_N = 9$~K. The structure turns around 3~K into another incommensurate one, which is described by two different
propagation vectors, namely, $\boldsymbol{k}_2=[\frac{5}{6},0.16,\frac{1}{2}]$ and $3\boldsymbol{k}_2$.
Below 2~K the structure related to $\boldsymbol{k}_1$ reappears and a coexistence of both above mentioned
magnetic structures is observed. Heat capacity data reveal that the transition at 3~K is of the first-order type.

%\section*{Declaration of Competing Interest}

%The authors declare no conflict of interest.

\section*{Acknowledgements}

%The research was partially carried out with the equipment purchased thanks to the financial
%support of the European Regional Development Fund in the framework of the Polish Innovation
%Economy Operational Program (contract no. POIG.02.01.00-12-023/08).

Kind hospitality and financial support extended to two of us (S.~B. and A.~S.) by the
Helmholtz-Zentrum Berlin f\"ur Materialien und Energie (HZB) is gratefully acknowledged.

%The open-access publication of this article was funded by the Priority Research Area SciMat under the program
%``Excellence Initiative – Research University'' at the Jagiellonian University in Krakow.



%%\section*{References}

%\bibliographystyle{elsarticle-harv} % author year scheme
%\bibliographystyle{elsarticle-num} % numbered scheme
%\bibliographystyle{elsarticle-num-names} %  numbered with new options of natbib.sty
%\bibliography{R2Ni2In}

%\begin{thebibliography}{00}


\end{document}